\documentclass[conference]{IEEEtran}
\usepackage{color, soul}
\usepackage{cite}
\usepackage{amsmath,amssymb,amsfonts}
\usepackage{algorithmic}
\usepackage{graphicx}
\usepackage{textcomp}
\usepackage{xcolor}
\usepackage[hyphens]{url}
\newcounter{rlabelno}

\usepackage{hyperref}
\hypersetup{%
    unicode=true,%
    pdfstartview={FitH},%
    colorlinks=true,%
}
\expandafter\def\expandafter\normalsize\expandafter{%
    \normalsize%
    \setlength\abovedisplayskip{3pt}%
    \setlength\belowdisplayskip{3pt}%
    \setlength\abovedisplayshortskip{3pt}%
    \setlength\belowdisplayshortskip{3pt}%
}

\def\BibTeX{{\rm B\kern-.05em{\sc i\kern-.025em b}\kern-.08em
    T\kern-.1667em\lower.7ex\hbox{E}\kern-.125emX}}

\title{A SAT Scalpel for Lattice Surgery: \\
{Representation and Synthesis of Subroutines for Surface-Code Fault-Tolerant Quantum Computing}}
\author{
\IEEEauthorblockN{Daniel~Bochen~Tan\IEEEauthorrefmark{1}\IEEEauthorrefmark{2}, Murphy~Yuezhen~Niu\IEEEauthorrefmark{3}\IEEEauthorrefmark{1}, Craig~Gidney\IEEEauthorrefmark{1}}
\IEEEauthorblockA{\IEEEauthorrefmark{1}Google Quantum AI\ \ \IEEEauthorrefmark{2}University of California, Los Angeles\ \ \IEEEauthorrefmark{3}University of California, Santa Barbara}
}

\begin{document}

\maketitle

%%%%%% -- PAPER CONTENT STARTS-- %%%%%%%%

\begin{abstract}

Quantum error correction is necessary for large-scale quantum computing.
A promising quantum error correcting code is the surface code.
For this code, fault-tolerant quantum computing (FTQC) can be performed via lattice surgery, i.e., splitting and merging patches of code.
Given the frequent use of certain lattice-surgery subroutines (LaS), it becomes crucial to optimize their design in order to minimize the overall spacetime volume of FTQC.
In this study, we define the variables to represent LaS and the constraints on these variables.
Leveraging this formulation,  we develop a synthesizer for LaS, LaSsynth, that encodes a LaS construction problem into a SAT instance, subsequently querying SAT solvers for a solution.
Starting from a baseline design, we can gradually invoke the solver with shrinking spacetime volume to derive more compact designs.
Due to our foundational formulation and the use of SAT solvers, LaSsynth can exhaustively explore the design space, yielding optimal designs in volume.
For example, it achieves 8\% and 18\% volume reduction respectively over two states-of-the-art human designs for the 15-to-1 T-factory, a bottleneck in FTQC.
\end{abstract}
\section{Introduction}\label{sec:intro}

%%% what is quantum computing and QEC
Quantum computing received broad interest because it has the potential to outperform conventional computing on certain problems.
For example, quantum computers are expected to break the public key cryptosystems that currently secure many internet transactions~\cite{shor-review}.
Quantum algorithms use millions or trillions of operations, but state-of-the-art quantum computers have gate errors rates ranging from one error per hundreds of gates to one error per ten thousand gates~\cite{google2023suppressing, moses2023race}.
This gap can be crossed using quantum error correction (QEC)~\cite{calderbank1998quantum, dennis2002topological, gottesman1997stabilizer}.

\begin{figure}[t]
    \centering
    \includegraphics[width=\linewidth]{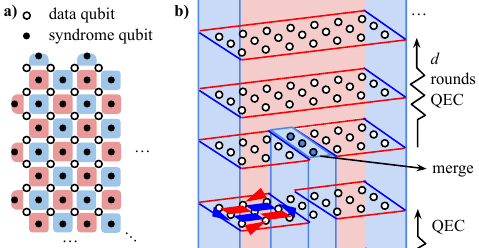}
    \vspace{-20pt}
    \caption{
    Fault-tolerant quantum computation based on the surface code.
    \textbf{a)}~The layout of physical qubits.
    Red faces stand for four-body or two-body $X$ stabilizers; blue faces are $Z$ stabilizers.
    We shall use color mapping ($X$,$Y$,$Z$) $\mapsto$ (R,G,B) in this paper.
    \textbf{b)}~An example logical operation: merging two separate tiles to a rectangular patch.
    The stabilizers (faces in a) for the left tile in the beginning are drawn.
    $d$ QEC rounds are performed after the merge.
    (Some rounds are omitted.)
    In the 3D spacetime, the boundaries of code patches (on the qubit plane) sweeping through time produce ``pipes''.
    }
    \label{fig:surface-code}
    % \vspace{-15pt}
\end{figure}

%%% what is surface code
Among the various QEC codes proposed, the surface code has emerged as a promising candidate for realizing large-scale fault-tolerant quantum computation (FTQC)~\cite{defect-surface-code, Litinski2019gameofsurfacecodes, litinski2022active}.
In this code, physical qubits are laid out as illustrated in \autoref{fig:surface-code}a.
Only nearest-neighbor connections are required.
There are two types of qubits: data qubits to encode quantum information, and syndrome qubits to detect errors.
In every QEC \textit{round}, each syndrome qubit measures a \textit{stabilizer} which is the parity of the four neighbor data qubits (or fewer if on boundaries) along the $X$ or $Z$ basis.
With these measurements, errors can be inferred, tracked, and corrected in the classical control software.

%%% what is lattice surgery
To perform FTQC, we need a scheme of computation acting on top of the surface code.
We opt for \textit{lattice surgery}~\cite{fowler2019low, lattice-surgery-first, Litinski2019gameofsurfacecodes} because it has a much lower resource overhead compared to alternative schemes like braiding~\cite{defect-surface-code, Raussendorf_2007}.
In lattice surgery, logical qubits are defined on \textit{patches} of surface code.
The simplest kind of patch is a \textit{tile}, e.g., the two tiles at the bottom of \autoref{fig:surface-code}b.
A tile has two types of boundaries, $X$ (red) or $Z$ (blue), predicated by the type of 2-body stabilizer on that boundary.
Tiles can be merged to larger patches.
Reversely, patches can be split to smaller patches.
The number of physical qubits in a tile is $d\times d$, where $d$ is the code distance which is the length of the longest error chain that can be caught.
Since the merging and splitting in lattice surgery only concern the boundaries of tiles, we \textit{treat a tile as the basic unit in space} in this work, independent from the code distance.
Similarly, $d$ QEC rounds are needed after every layer of operations, so we treat $d$ QEC rounds as one unit of time.
When the boundaries of patches sweep through time, the FTQC procedure becomes ``pipes'' in the 3D spacetime, exemplified by \autoref{fig:surface-code}b.

%%% compare with the scope of previous compilers
In contrast to previous compilation frameworks from quantum algorithms to lattice surgery~\cite{beverland2022assessing, temporal-lattice-surgery, fowler2019low, Litinski2019gameofsurfacecodes, litinski2022active, paler2020opensurgery, watkins2023high}, our focus is on hyper-optimizing small and frequently-used lattice-surgery subroutines (LaS).
Our tool is well suited for optimizing basic components with 5-20 qubits and 10-100 operations like the MAJ and UMA gates in adder circuits \cite{cuccaro2004adder}, or magic state distillation factories.
The intent is that, when assembling larger computations, the hyper-optimized subroutines created by our tool can be used as building blocks.
This is why, in the title, we analogize our tool to a \textit{scalpel}, not a \textit{Swiss army knife}.
Another major difference with previous approaches is the removal of human heuristics, aiming for provable optimality.
Unlike human constructions guided by some overarching ideas, our tool can explore results that are better but also highly non-intuitive.
The output usually comprises intricate arrangements of pipes and junctions, challenging to construct or even comprehend with human intuition.

\begin{figure}[t]
    \centering
    \includegraphics[width=\linewidth]{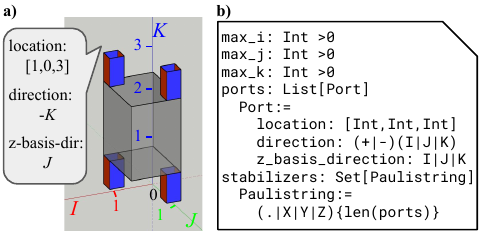}
    \vspace{-20pt}
    \caption{Lattice-surgery subroutine (LaS).
    \textbf{a)}~Externals of a LaS implementing a CNOT with four ports and $2\times 2 \times 2$ volume.
    \textbf{b)}~LaS specification including volume, port layout, and functionality (stabilizer) information.
    }
    \label{fig:subroutine}
    % \vspace{-5pt}
\end{figure}

\autoref{fig:subroutine} provides the specification of LaS.
The underlying surface code substrate is divided into tiles, our unit of space.
The tiles are indexed by spatial ($I$ and $J$ axes) coordinates, and they can only interact with nearest neighbors.
Similar to a classical circuit component, a LaS takes up a certain area on the quantum chip, i.e., \textit{footprint}.
In the example of \autoref{fig:subroutine}a, we allow $2 \times 2$ tiles: (0,0), (0,1), (1,0), and (1,1) in the $I$-$J$ plane.
A LaS is a procedure instead of a static entity, so it also has a duration (along $K$ axis).
In \autoref{fig:subroutine}a, we allow two units of time: $k=1$ and 2, each corresponds to a layer of operations and $d$ rounds of QEC afterwards.
Aligning with existing literature, we use \textit{spacetime volume}, i.e., the footprint times the duration, as our metric for LaS optimization.
The volume is indicated by the half-transparent box in our example.
Inside this box are the operations implementing the function of the LaS.
Note that the LaS specification does \textit{not} contain what happens inside this box.
Rather, it is the job of a synthesizer or compiler to figure out a solution satisfying the given specification.

How can we specify the function of a LaS?
For a classical digital circuit, this can be done by a lookup table between input and output ports.
A LaS also has some \textit{ports} connecting the given volume to the outside, e.g., the 4 short pipes in \autoref{fig:subroutine}a.
We first need to locate these ports, as in the list of \texttt{Port} in \autoref{fig:subroutine}b.
The \texttt{location} of a port is the outside 3D grid point it connects to, e.g., $(i,j,k)=(1, 0, 3)$ for the called-out port in \autoref{fig:subroutine}a.
Then, the \texttt{direction} of the port is the direction from this outside point to the inside.
In our example, from $(1,0,3)$, we need to go down to enter the box, so the direction is $-K$.
Finally, because of the two types of boundaries (red/blue), orientation of a port is also important.
In our example, the blue ($Z$) boundary is perpendicular to the $J$ axis, so the value of \texttt{z-basis-dir} is $J$.
The four ports in \autoref{fig:subroutine}a are inputs (bottom) and outputs (top) for control (left) and target (right) qubits in a CNOT gate.
To express the function of this LaS, the \texttt{stabilizers} in \autoref{fig:subroutine}b are \textit{stabilizer flows} on its ports: $ZI\to ZI$, $IZ\to ZZ$, $XI\to XX$, and $IX\to IX$.\footnote{An $n$-qubit Clifford can be represented by a matrix (a transition from state vectors to state vectors), but also can be defined by 2$n$ stabilizer flows~\cite{stabilizer-tableau}, e.g., if $U^\dagger (Z\otimes I) U=Z\otimes I$, $U^\dagger (I\otimes Z) U=Z\otimes Z$, $U^\dagger (X\otimes I) U=X\otimes X$, and $U^\dagger (I\otimes X) U=I\otimes X$, then the 2-qubit $U$ must be a CNOT.}
Note that these stabilizer flows are on the logical level, not to be confused with stabilizers measured on the physical level (inside tiles) in QEC rounds.

\begin{figure}[t]
    \centering
    \includegraphics[width=\linewidth]{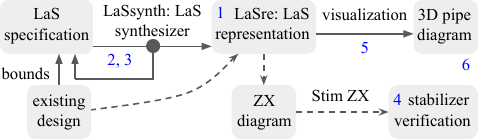}
    \vspace{-20pt}
    \caption{
    Overview of contributions (blue numbers).
    Solid arrows represent the main workflow.
    Dashes show the verification workflow for existing designs.
    }
    \label{fig:overview}
    % \vspace{-15pt}
\end{figure}

Our main task is to generate valid LaS given a specification above covering allowed volume, port configurations, and stabilizers to satisfy.
An overview of this paper is provided in \autoref{fig:overview}.
The main contributions are identified as follows.

\textcolor{blue}{1 Representation.}
Previous FTQC compilation works have introduced instruction sets atop lattice surgery, such as multi-qubit $\pi/8$-rotations and measurements~\cite{Litinski2019gameofsurfacecodes}, or a more general `planar quantum ISA'~\cite{beverland2022assessing}.
A main effort is decomposing arbitrary quantum algorithms into these instruction sets, which then straightforwardly unroll to known lattice surgery implementations.
While this abstraction layer of instruction set is necessary for efficiently compiling large-scale algorithms, it falls short when optimizing limited-scale LaS, as it cannot explore all potential combinations of lattice surgery.
In this work, we present the first LaS representation, LaSre, at the \textit{native} lattice surgery level, such that merging and splitting can potentially happen between any adjacent patches in the plane at any time, thus breaking the previous abstraction layer.

\textcolor{blue}{2 Formulation.}
Some \textit{validity constraints} are required to ensure that LaS expressed in our representation are valid.
For instance, two adjacent patches can only merge by the same type of boundary.
In the merge illustrated by \autoref{fig:surface-code}b, this constraint means the two pipes need to have the same color of boundaries (blue) facing each other.
Beyond validity, we also need to impose some \textit{functionality constraints} such that the resulting LaS realizes the stabilizer flows in the specification.
This is done by keeping track of objects named \textit{correlation surfaces}~\cite{Raussendorf_2007, Paler2015} (detailed later).
In summary, our formulation establishes a comprehensive set of constraints for the LaS synthesis problem based on our representation.

\textcolor{blue}{3 Synthesizer.}
Based on our formulation, we build LaSsynth, a software that transforms LaS synthesis into satisfiability (SAT) and queries SAT solvers for solutions.
Given a LaS specification like \autoref{fig:subroutine}b, the synthesizer either concludes that it is impossible to implement (UNSAT) or produces a satisfying variable assignment.
Existing designs provide us with concrete upper bounds in space and time to implement certain LaS.
In order for further optimization, we can start from these bounds, and iteratively generate and solve LaS specifications with even lower bounds, indicated by solid arrows in \autoref{fig:overview}.
Given the internal use of SAT by the synthesizer, its primary use case is optimizing frequently used subroutines, which can be pre-derived and integrated with end-to-end FTQC compilers.
The evaluations in this paper demonstrate that LaSsynth can solve subroutines of practical significance within a reasonable time frame.

\textcolor{blue}{4 Verification.}
We provide a verification workflow utilizing ZX calculus, as illustrated by the dashes in \autoref{fig:overview}.
Our formulation ensures correctness automatically, so the results generated by LaSsynth do not need to go through this.
However, when we were developing this software, the verification is very helpful in debugging.
Additionally, we can use it to check LaS designed by others, e.g., we found that the majority gate in Ref.~\cite{gidney2019flexible} does not realize some required stabilizer flows, underscoring the importance of automatic verification.

\begin{figure*}[t]
    \centering
    \includegraphics[width=\linewidth]{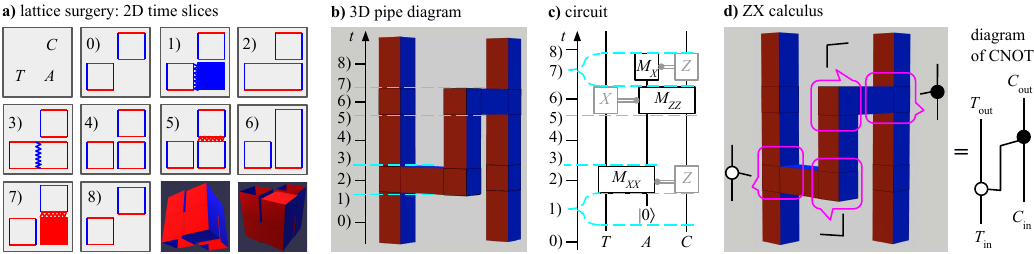}
    \vspace{-20pt}
    \caption{Logical CNOT in different representations.
    \textbf{a)}~2D time slices for lattice surgery.
    The floorplan is in the upper-left corner ($C$: control, $T$: target, $A$: ancilla).
    1-3) show the first merge-split.
    5-7) show the second merge-split.
    0), 4), and 8) are the states between these operations.
    \textbf{b)}~3D pipe diagram for lattice surgery.
    We mark the time slices in a) on the time axis.
    \textbf{c)}~CNOT quantum circuit using parity measurements.
    Slice 1) includes the initialization.
    Slice 7) includes the $X$ measurement.
    \textbf{d)}~ZX calculus.
    Junctions of pipes correspond to ZX spiders.
    The whole pipe diagram maps to the ZX diagram of a CNOT.
    }
    \label{fig:cnot}
    % \vspace{-10pt}
\end{figure*}

\textcolor{blue}{5 Visualization.} 
Until now, researchers have to manually construct the pipes in professional 3D modeling software~\cite{gidney2019flexible, fowler2019low}; or reply on 2D time slices like \autoref{fig:cnot}a \cite{Litinski2019gameofsurfacecodes, paler2020opensurgery}.
These compromises hinder lattice surgery research.
In response, we have developed a translation script from our representation to a 3D modeling format, facilitating visualization of LaS.

\textcolor{blue}{6 Specific designs.}
We discovered a majority gate design—a frequently used subroutine in Shor's algorithm—with a 40\% reduction in volume compared to previous work~\cite{gidney2019flexible}.
We also applied LaSsynth to optimize 15-to-1 $T$-factories.
Leveraging our foundational formulation and exhaustive search, it improves two state-of-the-art designs by 8\% than Ref.\cite{fowler2019low, Gidney2019efficientmagicstate} and by 18\% than Ref.\cite{Litinski2019gameofsurfacecodes}, under their respective settings.
Since $T$-factories can occupy 30\% of the spacetime volume in Shor's algorithm~\cite{Gidney2021howtofactorbit}, our result potentially reduces resource requirement of Shor's algorithm by 30\%$\times$18\%$=$5.4\%.

The paper is organized as follows.
In \autoref{sec:background}, we cover the background on lattice surgery.
\autoref{sec:formulation} presents the formulation of the LaS synthesis problem.
In \autoref{sec:implementation}, we provide the details on the implementation of LaSsynth.
In \autoref{sec:evaluation}, we apply LaSsynth to graph state generation, majority gate, and $T$-factory.
In \autoref{sec:related}, we survey previous works.
Finally, in \autoref{sec:conclusion}, we conclude the paper and provide future directions.

\begin{figure}[t]
    \centering
    \includegraphics[width=\linewidth]{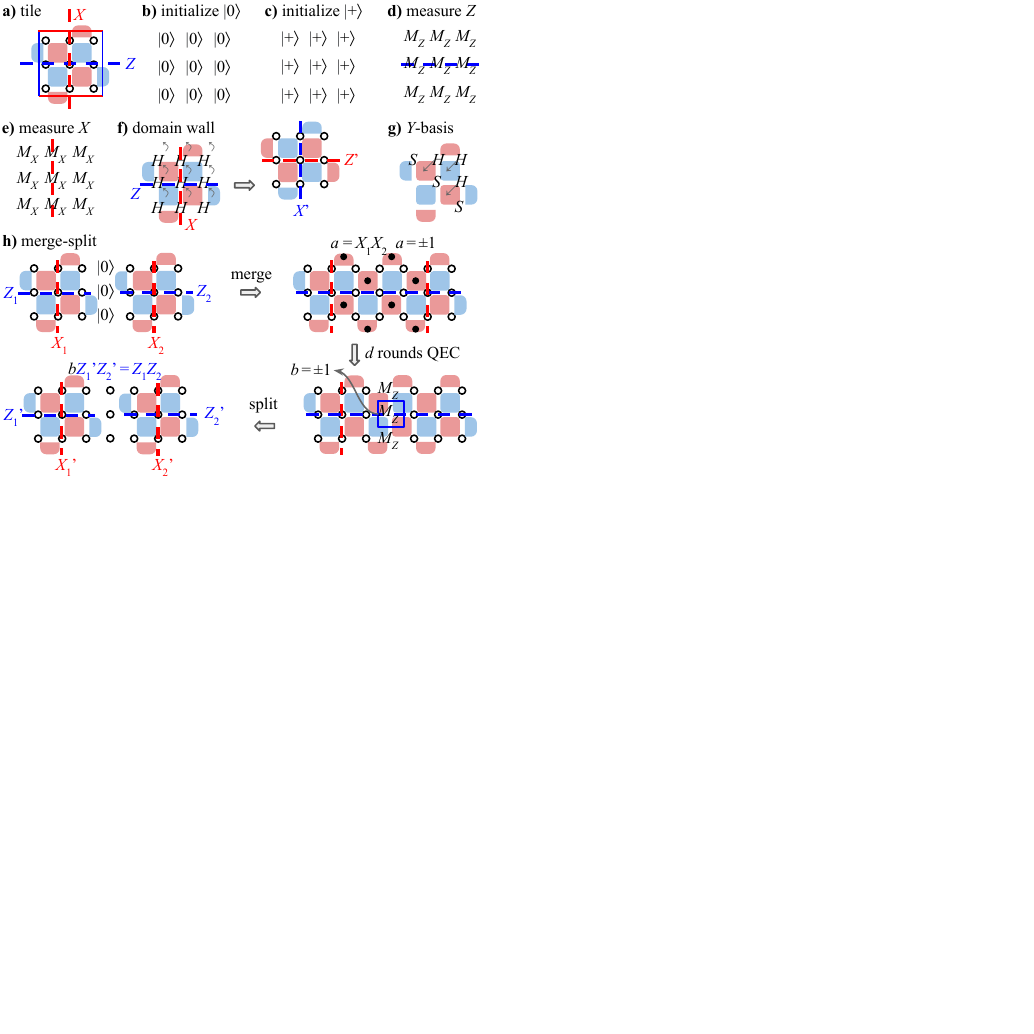}
    \vspace{-20pt}
    \caption{Surface code operations.
    \textbf{a)}~Surface code tile (simplest patch).
    \textbf{b-g)}~Single-patch operations.
    \textbf{h)}~Multiple-patch operation with lattice surgery.}
    \label{fig:lattice-ops}
    % \vspace{-15pt}
    % updated
\end{figure}

\section{Background} \label{sec:background}
We cover background knowledge on four topics: surface code operations which are the building blocks of LaS; the pipe diagram which is our graphical representation of LaS; the correlation surface, which is how we check what function a LaS implements; and the connection between lattice surgery and ZX calculus, which is how we can verify LaS.

\subsection{Surface Code Operations Considered} \label{ssec:surface-ops}
The basic unit in our FTQC scheme is a tile of surface code, e.g., in \autoref{fig:lattice-ops}a.
A tile has four boundaries of either $X$ or $Z$ type, indicated by red and blue solid lines, respectively.
Two opposite boundaries are of the same type when the tile is not being initialized or measured.
The logical $X$ operator is the tensor product of physical $X$ on a string of data qubits connecting two $X$ boundaries as indicated by the red dashes.
Similarly, the blue dashes indicate logical $Z$.
\autoref{fig:lattice-ops}b-c cover the initialization of logical $|0\rangle$ and $|+\rangle$, which are simply initializing all data qubits to $|0\rangle$ and $|+\rangle$, respectively.
\autoref{fig:lattice-ops}d-e cover the measurement of logical $X$ and $Z$, which are simply measuring all data qubits along the $X$ and $Z$ basis, respectively.
The logical measurement results are the products of measurements on the red or blue dashes, after error correction.
A key result of the above definition is that single-qubit Pauli gates can be applied \textit{off-chip} (in the classical control system) by interpreting measured data differently, e.g., if there is a logical $Z$ gate right before a logical $X$ measurement, we can just measure $X$ and flip our result, instead of actually applying the $Z$ gate \textit{on-chip} with a string of physical $Z$ gates.

\autoref{fig:lattice-ops}f introduces an operation named \textit{domain wall} realized by applying a layer of physical Hadamard gates and then some swap gates to shift data qubits (angled arrows).
It is useful when we need to rotate tiles by 90 degrees to a desirable orientation to interact with other patches.
\autoref{fig:lattice-ops}g corresponds to initialization or measurement along the $Y$ basis following Ref.~\cite{gidney2023inplace}.
The specific protocol to realize this operation consists of $d/2$ layers of physical gates where $d$ is the code distance.
Since we treat such protocol as an atomic operation in this work, we shall not dive into the details of it.

Lattice surgery operations consist of merge and split on $X$ or $Z$ boundaries~\cite{zx-lattice-surgery, lattice-surgery-first}.
When tiles merge, they can become non-square patches.
In this work, we employ a composite operation \textit{merge-split}: merge patches, perform $d$ rounds of QEC, and then split patches~\cite{fowler2019low}.
This merge-split definition is slightly restricted in the sense that a split does not have to be \textit{immediate} after the $d$ rounds of QEC.
However, we believe that keeping the merged patch does not offer additional benefits.
So, we follow this merge-split definition from Ref.~\cite{fowler2019low}.

\autoref{fig:lattice-ops}h presents the simplest merge-split on $Z$ boundaries of two patches.
In the merge, we initialize the data qubits in the ``gap'' between two patches to $|0\rangle$ and turn on all the syndrome qubits on the boundaries.
As a result, the logical operator $X_1X_2$ is measured in $a$, the product of the syndrome qubit measurements indicated by the solid dots.
The new $Z$ operator (long blue dashes) is the product of $Z_1$ and $Z_2$.
After $d$ QEC rounds, we perform the split where the data qubits in the gap are measured in the $Z$ basis.
Suppose the measurement in the box yields result $b$, we have $bZ'_1Z'_2=Z_1Z_2$.
This measurement does not touch $X$ operators, so $X'_1X'_2=X_1X_2=a$.
In summary, a merge-split of two tiles on $Z$ boundaries implements a logical $XX$ measurement; similarly, a merge-split on $X$ boundaries implements a $ZZ$ measurement.

\subsection{Pipe Diagram for LaS (Lattice-Surgery Subroutine)} \label{ssec:pipe}
To reason about the logical operations introduced above, we do not need all the information in \autoref{fig:lattice-ops}.
For instance, the code distance, $d$, is decided based on fidelity of physical gates and the total error budget and is independent from the computation carried out in a LaS.
To represent the on-chip process in a distance-independent manner, we can use time slices such as \autoref{fig:cnot}a.
These operations implement a CNOT between the control tile ($C$) and the target tile ($T$) via an ancillary tile ($A$). 
Each tile corresponds to $d \times d$ data qubits physically, but we only need to keep its boundaries in the representation.
Slices 1-4) correspond exactly to the merge-split exhibited in \autoref{fig:lattice-ops}h, while slices 4-7) correspond to a merge-split on $X$ boundaries.
The color-filled tiles correspond to logical initializations or measurements as \autoref{fig:lattice-ops}b-e.

Why does this sequence of operations implement a CNOT?
We can compare the quantum circuits of a CNOT using parity measurements~\cite{fowler2019low}, \autoref{fig:cnot}c, with the time slices.
Slice 1) simultaneously initializes $A$ to $|0\rangle$ and starts a merge-split on $Z$ boundaries of $A$ and $T$, i.e., the $XX$ measurement of $A$ and $T$.
Slice 2) is during the $d$ rounds of QEC in the merge-split, and slice 3) is finishing the merge-split.
Note that the gray components in the circuit are Pauli gates depending on the measurement results.
They do not appear in the slices because they are off-chip.
Slices 5-7) correspond to the second merge-split between $A$ and $C$ and also the $X$ measurement of $A$.

The time slices are ``snapshots'' of what happens on the quantum chip at important moments.
Many moments during QEC are neglected because they would be the same with previous moments.
To also capture the time dimension in the representation, we introduce the 3D pipe diagram.
A tile of distance $d$ needs $d$ rounds of QEC after an operation, so its boundaries ``stay put'' for $d$ rounds.
When we trace these boundaries on a time axis, a tile accumulates to a cube.
The two small 3D drawings in \autoref{fig:cnot}a exhibit such \textit{in-scale} pipe diagrams for the CNOT.
However, the merge-splits in the in-scale drawings are in the narrow gaps between cubes, not very visible, so we elongate the distance between the cubes so that the merge-splits become pipes as shown in \autoref{fig:cnot}b.
The 3D pipe diagram is simpler and more intuitive than the slices, e.g., a merge-split is just a horizontal pipe instead of multiple slices.
The continuations and terminations of vertical pipes indicates initializations and measurements.
For example, the middle pipe terminates at 7) when the ancilla is measured out. 

\subsection{Correlation Surfaces} \label{ssec:correlation-surfaces}

We have just proved that \autoref{fig:cnot}b implements a CNOT by establishing equivalence with a circuit.
However, it is hard to construct LaS entirely from circuits because 1) lattice surgery is non-unitary, different from the unitary gates we are used to; and 2) some critical information are missing, like the layout and orientation of tiles which determine whether a merge-split is valid.
In this paper, we ensure the correctness of LaS with objects called correlation surfaces that travel inside the pipes.

\begin{figure}[t]
    \centering
    \includegraphics[width=\linewidth]{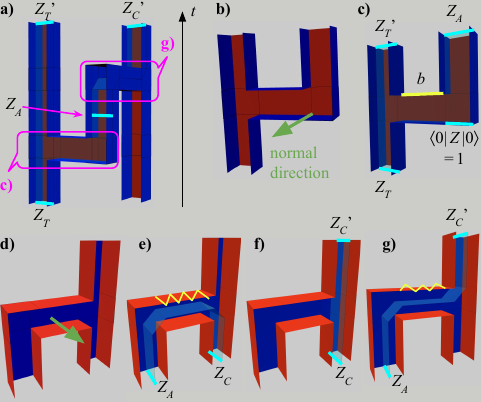}
    \vspace{-20pt}
    \caption{
    Examples of correlation surfaces.\ \  
    \textbf{a)}~A surface that ensures the stabilizer flow $Z_T\to Z'_T Z'_C$ in a CNOT pipe diagram (front face removed).
    It is a product of two local pieces in c) and g).
    \textbf{b-c)}~Two cases where the surface is orthogonal to the normal direction.
    \textbf{d-g)}~Four cases where the surface is parallel to the normal direction.
    The sign of the correlated operators is the product of measurements on either a line of qubits, highlighted in c), or a rectangular region of qubits, indicated by the wiggled lines in e) and g).
    }
    \label{fig:corrsurf}
    % \vspace{-15pt}
\end{figure}

In \autoref{fig:corrsurf}a, we exhibit a correlation surface ensuring the LaS satisfies a stabilizer flow $Z_T \to Z'_C Z'_T$, one of the flows that define a CNOT.
The correlation surfaces for the other three flows should also be identified to ensure the correctness of this LaS.
These seemingly complicated surfaces are composed with simple local pieces at each junction of pipes, e.g., \autoref{fig:corrsurf}a is composed by \autoref{fig:corrsurf}c and g (by stitching over $Z_A$).

A correlation surface relates a set of quantum logical operators and a set of measurement results.
The operators are at all the ports this surface propagates to, and the measurements are the projections of this surface on the ceiling of all pipes it propagates in.
In \autoref{fig:corrsurf}c, the operators are $Z_T$, $Z'_T$, and $Z_A$ (highlighted in cyan), and the measurements are on the highlighted yellow line.
`Correlation' just means the product of these logical operators are equal to the product of the measurements.
So, why is this local piece valid?
Recall that the horizontal pipe in \autoref{fig:corrsurf}c is a merge-split on $Z$ boundaries as in  \autoref{fig:lattice-ops}h.
Since the ancilla is initialized to $|0\rangle$, its $Z$ value before the merge-split is fixed, $\langle 0 | Z | 0\rangle = 1$.
Thus, according to \autoref{fig:lattice-ops}h, $bZ'_T Z_A = Z_T\cdot 1$, i.e., $Z'_TZ_AZ_T=b$.
This is consistent with our definition since the left hand side is the product of all logical operators the surface touches, and the right hand side is the measurement on the projection of the surface to the ceiling of the horizontal pipe.
Note that we have elongated the horizontal pipes in the 3D diagram.
The highlighted line corresponds to only one measurement physically, which is the one producing $b$ in \autoref{fig:lattice-ops}h.
In real experiments, we will need the correlation surfaces to tell us which set of measurements decides the sign of these product of operators.
If we require $Z_TZ'_TZ_A=1$ while $b$ measures $-1$, some logical single-qubit Pauli gates should be applied off-chip to fix it.

No surface at all, e.g., \autoref{fig:corrsurf}b, is also valid.
In general, the \textit{rules} of correlation surfaces at junctions can be reasoned from the logical effects of merge-splits.
There are two categories: whether the surface is orthogonal to the \textit{normal direction} of the junction, or parallel to it.
The normal direction is orthogonal to all the pipes in the junction, as indicated by the green arrows in \autoref{fig:corrsurf}b and d.
If the surface is orthogonal to it, the surface correlates all (\autoref{fig:corrsurf}c) or none (\autoref{fig:corrsurf}b) of the logical operators at ports.
If the surface is parallel to it, the surface correlates an even number of logical operators at ports: in our example, there can only be 0 (\autoref{fig:corrsurf}d) or 2 operators (\autoref{fig:corrsurf}e-g).
The wiggly lines in \autoref{fig:corrsurf}e and g indicate the projections of correlation surfaces to the ceilings of pipes, which correspond to the parity measurement result $a$ in \autoref{fig:lattice-ops}h.
(Although this particular horizontal pipe is a $ZZ$ measurement instead of an $XX$ measurement.)
We will formally present all the rules for correlation surfaces in \autoref{ssec:functional-constraints}.
If at every junction, these rules are respected, the whole correlation surface is valid.

\subsection{Lattice Surgery and ZX Calculus} \label{ssec:lattice-surgery-zx}

\begin{figure}[t]
    \centering
    \includegraphics[width=\linewidth]{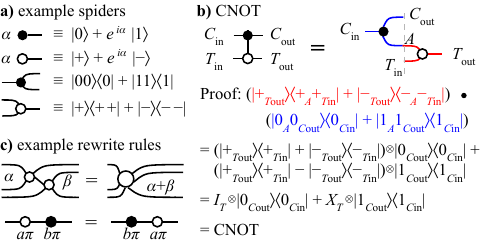}
    \vspace{-20pt}
    \caption{
    ZX calculus.
    \textbf{a)}~Examples of spiders.
    \textbf{b)}~Proof of the CNOT diagram.
    \textbf{c)}~Examples of rewrite rules: merge phases, and exchange whole-$\pi$ spiders.
    }
    \label{fig:zxintro}
    % \vspace{-15pt}
\end{figure}

ZX calculus is a graphical language with which we can intuitively perform algebra related to quantum computing.
We will briefly introduce ZX calculus in the context of this paper.
Interested readers can refer to Ref.~\cite{vandewetering2020zxcalculus} for more details.

As exhibited in \autoref{fig:zxintro}a, the basic components in ZX diagrams are two types of spiders, $Z$-spider (solid dots) and $X$-spider (circle).
These spiders are connected by wires.
(Our color scheme may deviate from existing literatures: $Z$-spiders and $X$-spiders are green/light and red/dark, respectively, in Ref.~\cite{zx-lattice-surgery} and \cite{vandewetering2020zxcalculus}.)
Each spider has a parameter named \textit{phase}, which is an angle in $[0,2\pi)$.
If the phase is not annotated, it is assumed 0, like the lower two spiders in \autoref{fig:zxintro}a.

\begin{figure*}[t]
    \centering
    \includegraphics[width=\linewidth, height=.3\linewidth]{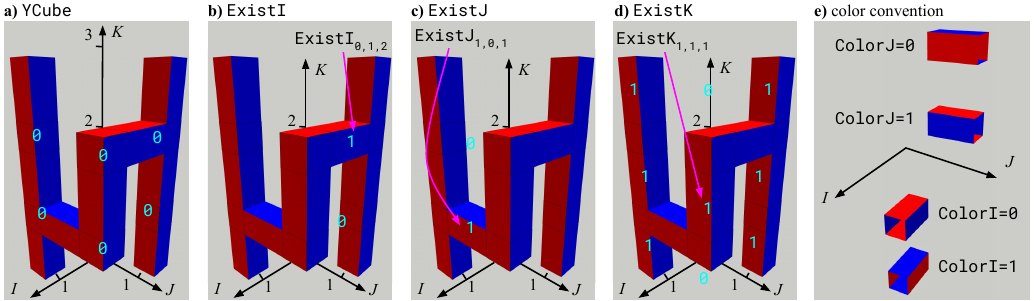}
    \vspace{-20pt}
    \caption{Structural variables.
    \textbf{a)}~\texttt{YCube} on each cube specifies whether that cube is a $Y$ cube.
    \textbf{b-d)}~\texttt{ExistI/J/K} on each edge specifies whether there is an edge in the corresponding direction.
    \textbf{e)}~color convention for \texttt{ColorI/J} variables.
    }
    \label{fig:structural-vars}
    % \vspace{-10pt}
\end{figure*}

Fundamentally, each spider is a tensor, and a wire between two spiders is a tensor contraction.
Thus, what matter are the labeling of open wires, the type and phase of spiders, and their connectivity.
In contrast, the geometric locations of spiders are irrelevant, and the wires can bend and stretch.
In \autoref{fig:zxintro}b, two equivalent ZX diagrams of CNOT are displayed, up to these meaning-preserving deformations.
We can apply the definition of spiders in the diagram on the right to prove it is indeed a CNOT.
The first step is due to the contraction at point $A$: $\langle \pm|0 \rangle=1$ and $\langle \pm|1 \rangle=\pm 1$.
(We ignore the constant $1/\sqrt 2$ here.)
The second step is by the definition of Pauli gates: $I=|+\rangle \langle+|+|-\rangle \langle-|$ and $X=|+\rangle \langle+|-|-\rangle \langle-|$.
The final step is the definition of a CNOT: in case of $|0\rangle$ on the control, do nothing; in case of $|1\rangle$ on the control, apply $X$ to the target.

Based on the algebra of spiders, some rewrite rules are derived to help one manipulate ZX diagrams, e.g., merging same-type spiders, and exchanging whole-$\pi$ phase spiders ($a,b=0,1$), as in \autoref{fig:zxintro}c.
We refrain from more details about rewrite rules since they are irrelevant at the moment.

It turns out lattice surgery has a very natural correspondence to ZX diagrams~\cite{zx-lattice-surgery, stabilizer-zx, bombin2023unifying}.
As a result, we can derive the ZX diagram of a LaS in a straightforward way (\autoref{fig:cnot}d) \cite{gidney2019flexible}: every cube is a spider, and every pipe is a wire connecting two spiders.
If a cube is simply a 90-degree turn, it is a wire.
For a T-junction or a cross-junction, the color of the junction is the color that draws the T or the cross.
In \autoref{fig:cnot}d, there is a red T-junction on the lower left and a blue T-junction on the upper right.
Then, the spiders have corresponding types, e.g., the red T-junction is a $X$-spider whereas the blue T-junction is a $Z$-spider.
All the phases of the spiders are 0 except for $Y$-basis operations, to appear later on, which have phase $\pi/2$.
Wiring these spiders together in \autoref{fig:cnot}d, we find that the pipe diagram indeed implements the ZX diagram of a CNOT.

\section{Problem Formulation} \label{sec:formulation}
Given a LaS specification as in \autoref{fig:subroutine}b, we formulate the LaS synthesis problem into binary variables and constraints that must be satisfied.
These variables constitute LaSre, our LaS representation.
The constraints are of two types: \textit{validity} and \textit{functionality}.
The former ensures that the LaSre is a valid FTQC procedure, whereas the latter ensures that it satisfies the stabilizers specified, which is the function of the LaS.

From this point on, we will use $I$ and $J$ as two spatial axes, and $K$ as the time axis.
This notational choice is because letters $X$, $Y$, $Z$, and $T$ have other meanings in the context.

\subsection{Structural Variables}

In a pipe diagram, each cube can potentially connect to its nearest neighbors in the 3D spacetime.
Thus, to specify the structure of the pipes, all we need is one bit for every adjacent pair of cubes meaning whether there is a pipe or not.
These are the \texttt{Exist*} variables below. 
Additionally, each horizontal pipe has a color orientation (\autoref{fig:structural-vars}e), denoted by \texttt{Color*} variables below, that corresponds to whether the pipe is a merge-split on $Z$ or $X$ boundaries.
Once the configurations of all pipes are known, the faces of cubes can be inferred from all its incident pipes, except for one case--$Y$ cubes, which are represented by green boxes in this paper.
These correspond to initializing and measuring patches along the $Y$ basis (\autoref{fig:lattice-ops}g).

There are five arrays of structural variables.
Each has shape (\texttt{max\_i}, \texttt{max\_j}, \texttt{max\_k}), i.e., one binary variable per spacetime volume.
All indices start from 0.
\autoref{fig:structural-vars} shows the values of these variables in the CNOT example.

$\texttt{YCube}_{i,j,k}$ specifies whether the cube at $(i,j,k)$ is a $Y$ cube.
In this example, there is none, so all $\texttt{YCube}_{i,j,k}=0$.

$\texttt{ExistI}_{i,j,k}$ specifies whether there is a pipe from cube $(i,j,k)$ to $(i+1,j,k)$.
There is only one pipe in $I$ direction, which is from $(0,1,2)$ to $(1,1,2)$ (see \autoref{fig:structural-vars}b).
So, only $\texttt{ExistI}_{0,1,2}=1$, while all other $\texttt{ExistI}$ variables evaluates to 0, e.g., directly below the existing $I$-pipe, $\texttt{ExistI}_{0,1,1}=0$.
Note that our convention allows pipes from $(\texttt{max\_i}-1,*,*)$ to $(\texttt{max\_i},*,*)$, which extend out of the confined volume.
These can only be ports.
However, we do not allow -1 as an index, so there is no pipe from $(-1,*,*)$ to $(0,*,*)$.

$\texttt{ExistJ}_{i,j,k}$ specifies whether there is a pipe from cube $(i,j,k)$ to $(i,j+1,k)$.
In the example, there is only one $J$-pipe from $(1,0,1)$ to $(1,1,1)$, so $\texttt{ExistJ}_{1,0,1}=1$ (\autoref{fig:structural-vars}c), and all other $\texttt{ExistJ}$ variables are 0, e.g., $\texttt{ExistJ}_{1,0,2}=0$.

$\texttt{ExistK}_{i,j,k}$ specifies whether there is a pipe from cube $(i,j,k)$ to $(i,j,k+1)$.
There are many $K$-pipes in the example, indicated by the 1's in \autoref{fig:structural-vars}d.
Although the merge-splits only concerns horizontal pipes, we still need these variables for vertical pipes because they indicate initializations and measurements.
For example, $\texttt{ExistK}_{1,1,0}=0$, $\texttt{ExistK}_{1,1,1}=1$, and $\texttt{ExistK}_{1,1,2}=0$ means the tile at $(i,j)=(1,1)$ is initialized at $k=1$ and measured at $k=2$.

$\texttt{ColorI}_{i,j,k}$ specifies the color orientation of $I$-pipes.
Our convention is displayed in \autoref{fig:structural-vars}e: if red faces are on the $K$ direction, then $\texttt{ColorI}=0$; otherwise, $\texttt{ColorI}=1$.
The $\texttt{Color*}$ variables are always used together with $\texttt{Exist*}$ variables in the constraints, so if there is no pipe, the corresponding color variable value will not affect the solution. 

$\texttt{ColorJ}_{i,j,k}$ specifies the color orientation of $J$-pipes.

At this point, it seems there should also be $\texttt{ColorK}$ variables.
Indeed, in the 3D drawings, $K$-pipes are also colored.
However, we do not need $\texttt{ColorK}$ variables because the domain wall operation is available to us.
This operation is denoted as a yellow ring on a $K$-pipe, as found in \autoref{fig:graph-state}b (at the green `5').
It can switch the orientation in the middle of a $K$-pipe to adapt it to any configuration at the two ends.
Thus, $\texttt{ColorK}$ variables are neglected in the formulation, and inferred in the post-processing based on its neighbors.

\subsection{Validity Constraints} \label{ssec:structural-constraints}
\begin{figure}[t]
    \centering
    \includegraphics[width=\linewidth]{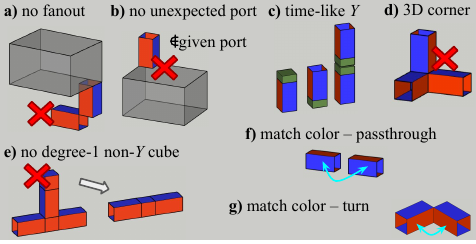}
    \vspace{-20pt}
    \caption{Validity constraints.
    \textbf{a)}~Starting point of a port cannot have other pipes.
    \textbf{b)}~No pipes other than the ports connect to the outside.
    \textbf{c)}~$Y$ cubes can only have $K$-pipes.
    \textbf{d)}~A cube cannot have pipes in all three directions.
    \textbf{e)}~All degree-1 non-$Y$ cubes can always ``squeeze in'', so we forbid such cubes.
    \textbf{f)}~Two pipes in a ``passthrough'' should have the same color orientation.
    \textbf{g)}~Two pipes in a ``turn'' should have matching colors on faces that are touching.
    }
    \label{fig:structural-constraints}
    % \vspace{-15pt}
\end{figure}

\autoref{fig:structural-constraints} illustrates the validity constraints.
Rules a) and b) follow from our definition of a port; c) is required for the $Y$-basis operation~\cite{gidney2023inplace}; d), f), and g) follow from the fact that split-merge can only happen when the adjacent boundaries of two patches have the same type.
Rule e) is technically an optimization in order for lower spacetime volume instead of a hard requirement.
We provide some examples below.

\textit{No Fanouts} (\autoref{fig:structural-constraints}a).
A port in the LaS specification starts at a cube and connects to \textit{only one} of its 6 neighbors.
For instance, port 0 in \autoref{fig:functional-vars} starts from $(0,1,0)$ and connects $(0,1,1)$, so there is no connection in the other 5 directions:
\begin{equation}
    \overline{\texttt{ExistI}_{0,1,0}}\ ,\ \ \overline{\texttt{ExistJ}_{0,0,0}}\ ,\ \ \overline{\texttt{ExistJ}_{0,1,0}}\ ,
\end{equation}
where an overline means the variable should be 0.
Note that the pipes in $-I$ and $-K$ directions for this cube are out of range, so they are automatically neglected.

\textit{No Unexpected Ports} (\autoref{fig:structural-constraints}b).
Other than the specified ports, there cannot be ``dangling'' pipes.
In \autoref{fig:functional-vars}, on the top floor, there are only port 2 starting at $(0,1,2)$ and port 3 at $(1,0,2)$, so the other two patches cannot connect upwards:
\begin{equation}
    \overline{\texttt{ExistK}_{0,0,2}}\ ,\ \ \overline{\texttt{ExistK}_{1,1,2}}\ .
\end{equation}

\textit{Time-Like $Y$ Cubes} (\autoref{fig:structural-constraints}c).
Only time-like ($K$) pipes are allowed to connect to a $Y$ cube, so
\begin{equation} \label{eq:time-line-y}
    \forall (i,j,k)\ \  \texttt{YCube}_{i,j,k}\ \Rightarrow\ \overline{\texttt{ExistI}_{i-1,j,k}}\ .
\end{equation}
We also apply similar constraints with $\overline{\texttt{ExistI}_{i,j,k}}$ , $\overline{\texttt{ExistJ}_{i,j-1,k}}$ , and $\overline{\texttt{ExistJ}_{i,j,k}}$ on the right hand side.

\textit{No Degree-1 Non-$Y$ Cubes} (\autoref{fig:structural-constraints}e).
Degree-1 cubes (those having only one pipe) that are neither $Y$-cubes nor ports can always be ``squeezed in''.
In \autoref{fig:structural-constraints}e, the degree-1 cube measures logical $Z$ of the patch, but that measurement can be on the horizontal ``passthrough'' without ``popping out''.
This constraint is expressed as follows: for an endpoint of a pipe, if it is not a $Y$ cube, then at least one of the other 5 pipes of this cube is present, e.g., for $K$-pipe $(1,0,1)$ to $(1,0,2)$,
\begin{equation} \label{eq:no-degree-1}
    \begin{split}
        &\overline{\texttt{YCube}_{1,0,1}} \wedge\texttt{ExistK}_{1,0,1}\ \Rightarrow\ \big[ \texttt{ExistK}_{1,0,0} \vee \\
        &\ \ \ \ \ \ \ \ \ \ \texttt{ExistI}_{1,0,1} \vee \texttt{ExistI}_{0,0,1} \vee \texttt{ExistJ}_{1,0,1} \big].
    \end{split}
\end{equation}
The right hand side has 4 terms instead of 5 terms because the pipe in $-J$ direction for cube $(1,0,1)$ is out of range.

\textit{Matching Colors at Passthroughs} (\autoref{fig:structural-constraints}f).
Two pipes in the same direction connecting to a cube should have the same color orientation, e.g., the two possible $I$ pipes of $(1,1,2)$:
\begin{equation}
    \begin{split}
         \big[ \texttt{ExistI}_{0,1,2}\ \wedge\ &\texttt{ExistI}_{1,1,2} \big]\Rightarrow \\
        & \big[ \texttt{ColorI}_{0,1,2} = \texttt{ColorI}_{1,1,2}\big],
    \end{split}
\end{equation}
which is satisfied (trivially) because $\texttt{ExistI}_{1,1,2}=0$.

\textit{Matching Colors at Turns} (\autoref{fig:structural-constraints}g).
When two pipes in orthogonal directions are touching, their faces that are touching should have the same color.
Consider two possible pipes connecting to $(1,1,2)$ in $-I$ and $-J$ directions:
\begin{equation}
    \begin{split}
        \big[ \texttt{ExistI}_{0,1,2}\ \wedge\ &\texttt{ExistJ}_{1,0,2} \big]\Rightarrow \\
        &\big[ \texttt{ColorI}_{0,1,2} \neq \texttt{ColorJ}_{1,0,2}\big],
    \end{split}
\end{equation}
which is satisfied because $\texttt{ExistJ}_{1,0,2}=0$.
However, suppose there is a pipe, we can check \autoref{fig:structural-vars}e that if the $\texttt{ColorI}$ is 0, to match the colors, the $\texttt{ColorJ}$ must be 1.
Alternatively, the former is 1 and the latter is 0.
These two cases are exactly captured by the right hand side with the Boolean $\neq$ operator.

\textit{No 3D Corners} (\autoref{fig:structural-constraints}d).
If a cube connects to at least one pipe in all $I$, $J$, and $K$ directions, it is a ``3D corner'', which is forbidden.
In \autoref{fig:structural-constraints}d, the $K$-pipe and the $J$-pipe have a color matching conflict.
Switching the orientation of the $K$-pipe, then it has a conflict with the $I$-pipe.
This set of constraint is formulated as each cube having a normal direction, i.e., it does not have pipes completely in at least one of the $I$, $J$, or $K$ direction.
Consider cube $(1,1,2)$ again:
\begin{equation} \label{eq:no-3d-corner}
\begin{split}
    & \big[ \  \overline{\texttt{ExistI}_{0,1,2}} \wedge \overline{\texttt{ExistI}_{1,1,2}} \ \big] \vee \big[ \ \overline{\texttt{ExistJ}_{1,0,2}}\  \wedge \\
    & \ \ \ \ \ \ \overline{\texttt{ExistJ}_{1,1,2}} \ \big] \vee \big[ \ \overline{\texttt{ExistK}_{1,1,1}} \wedge \overline{\texttt{ExistK}_{1,1,2}} \ \big],
\end{split}
\end{equation}
which is satisfied since $\texttt{ExistJ}_{1,0,2}=\texttt{ExistJ}_{1,1,2}=0$.

\subsection{Correlation Surface Variables}
\begin{figure}[t]
    \centering
    \includegraphics[width=\linewidth]{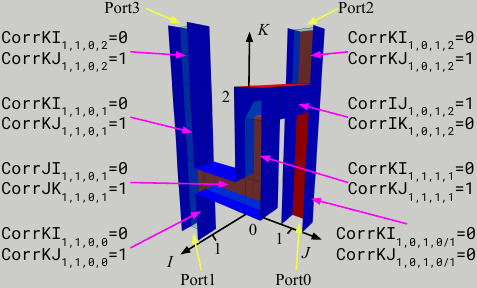}
    \vspace{-20pt}
    \caption{Correlation surface variables. The surface for $IZZZ$ is shown.}
    \label{fig:functional-vars}
%    \vspace{-10pt}
\end{figure}

For each stabilizer of a LaS, we need to provide the corresponding correlation surface which can be specified by its pieces inside each pipe.
Two pieces are possible inside a pipe: one that connects the two blue faces and one that connects the two red faces.
Thus, we need two bits per pipe for each of the correlation surfaces.
These are the \texttt{Corr**} variables below.
There are six arrays of correlation surface variables, each has shape ($n_\text{stab}$, \texttt{max\_i}, \texttt{max\_j}, \texttt{max\_k}), one binary variable per stabilizer per volume.
\autoref{fig:functional-vars} provides the correlation surface variable values in each pipe for the surface seen in \autoref{fig:corrsurf}a.

$\texttt{CorrIJ}_{s,i,j,k}$ specifies the existence of correlation surface pieces in the $I$-$J$ plane and inside $I$-pipes.
Note that $IZZZ$ has index $s=1$ among the 4 stabilizers, and there is only one $I$-pipe from $(0,1,2)$ to $(1,1,2)$.
Thus, $\texttt{CorrIJ}_{1,0,1,2}=1$.

$\texttt{CorrIK}_{s,i,j,k}$ is for correlation surface pieces in the $I$-$K$ plane and inside $I$-pipes.
Since there is only one $I$-pipe, the only nontrivial example here is $\texttt{CorrIK}_{1,0,1,2}=0$.

Similarly, we have variable arrays $\texttt{CorrJK}_{s,i,j,k}$ and $\texttt{CorrJI}_{s,i,j,k}$ for correlation surface pieces in $J$-pipes, and $\texttt{CorrKI}_{s,i,j,k}$ and $\texttt{CorrKJ}_{s,i,j,k}$ for those in $K$-pipes.
Since each stabilizer needs a different correlation surface, the number of correlation surface variables scales in the product of the volume and the number of stabilizers, much larger than the number of structural variables scaling only in the volume.

\subsection{Functionality Constraints} \label{ssec:functional-constraints}
\begin{figure}[t]
    \centering
    \includegraphics[width=\linewidth]{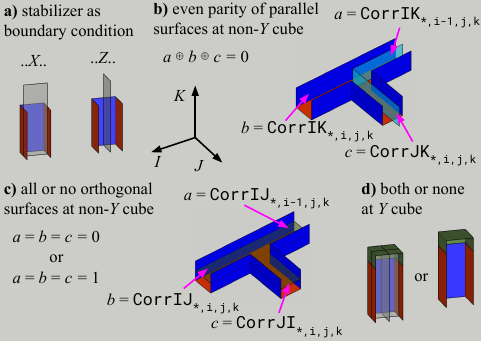}
    \vspace{-20pt}
    \caption{Functionality constraints.
    \textbf{a)}~At a port, the correlation surface should be consistent with the stabilizer: only connecting red faces if $X$; only connecting blue faces if $Z$; both if $Y$; neither if $I$.
    \textbf{b)}~An even number of the correlation surfaces parallel to the normal direction should be present.
    \textbf{c)}~All or none of the correlation surfaces orthogonal to the normal direction should be present.
    \textbf{d)}~Both or no correlation surfaces should be present at a $Y$ cube.
    }
    \label{fig:functional-constraints}
    % \vspace{-15pt}
\end{figure}

\autoref{fig:functional-constraints} illustrates the constraints for correlation surfaces: two at boundaries (a and d), and two at junctions (b and c).

\textit{Stabilizer as Boundary Conditions} (\autoref{fig:functional-constraints}a).
A stabilizer is specified as Pauli string at the ports, e.g., $IZZZ$ for the four ports of the CNOT.
The $Z$ operator touches the two $Z$ boundaries (blue) of the $K$-pipes of the ports.
Given that the \texttt{z-basis-dir} in the input is $J$ for all the ports, we find that there should be correlation surface pieces in the $J$-$K$ plane inside the $K$-pipes of ports 1, 2, and 3.
Also, the correlation surface pieces corresponding to $X$ operators, in the $K$-$I$ plane, should not be present at these ports.
Therefore,
\begin{equation}
    \begin{split}
        &\texttt{CorrKJ}_{1,1,0,0}\ ,\ \ \texttt{CorrKJ}_{1,0,1,2}\ ,\ \ \texttt{CorrKJ}_{1,1,0,2}\ , \\
        &\overline{\texttt{CorrKI}_{1,1,0,0}}\ ,\ \ \overline{\texttt{CorrKI}_{1,0,1,2}}\ ,\ \ \overline{\texttt{CorrKI}_{1,1,0,2}}\ .
    \end{split}
\end{equation}
Port 0 should not have correlation surface pieces in neither direction since in the stabilizer, its term is $I$.
If the term for a port is $Y$, then it should have both correlation surface pieces.

\textit{Both or None at $Y$ Cubes} (\autoref{fig:functional-constraints}d).
Since the $Y$ operator is the product of $Z$ and $X$, it correlates the two operators.
Thus, two correlation surfaces can end at a $Y$ cube together, or neither of them are present in this region.
Per \autoref{fig:structural-constraints}c, only $K$-pipes can connect to $Y$ cubes, so this set of constraint is
\begin{equation}
    \texttt{YCube}_{s,i,j,k}\ \Rightarrow\ \big[ \texttt{CorrKI}_{s,i,j,k} = \texttt{CorrKJ}_{s,i,j,k} \big],
\end{equation}
for all tuples of $(s,i,j,k)$.

A non-$Y$ and non-port cube can only have degree 2, 3, or 4: degree-1 cube is forbidden by \autoref{fig:structural-constraints}e; and degree-5 or -6 cubes will always contain a 3D corner which is forbidden by \autoref{fig:structural-constraints}d.
Degree-2 cubes are ``identity'' where correlation surfaces trivially travel through.
We can neglect them for now and check later that this case is consistent with the two sets of constraints to introduce below.
In conclusion, we are only left with degree-3 or -4 cubes which can only be T- or cross-junctions, and they each has a normal direction orthogonal to the plane that draws the T or the cross.
In \autoref{fig:functional-constraints}b-c, the T-junctions are in $I$-$J$ plane, so their normal direction is $K$.
The two constraints below depend on whether the correlation surfaces are orthogonal or parallel to the normal direction.

\textit{Even Parity of Parallel Surfaces at Non-$Y$ Cubes} (\autoref{fig:functional-constraints}b, generalizes \autoref{fig:corrsurf}d-g).
There should be an even number of correlation surface pieces parallel to the normal direction of the cube.
In \autoref{fig:functional-vars}, the normal direction for $(0,1,2)$ is $J$, so
\begin{equation}
    \begin{split}
        \big[ \ &\overline{\texttt{YCube}_{0,1,2}}\ \wedge \overline{\texttt{ExistJ}_{0,0,2}}\  \wedge \overline{\texttt{ExistJ}_{0,1,2}} \ \big] \Rightarrow \\
        & \big\{ \big[ \texttt{ExistI}_{0,1,2}\ \wedge \texttt{CorrIJ}_{1,0,1,2} \big] \oplus \big[ \texttt{ExistK}_{0,1,1} \ \wedge\\
        & \ \ \  \texttt{CorrKJ}_{1,0,1,1} \big] \oplus \big[ \texttt{ExistK}_{0,1,2} \wedge \texttt{CorrKJ}_{1,0,1,2} \big]  =0 \big\},
    \end{split}
\end{equation}
where the first and third term on the right hand side are 1 because the said correlation surface pieces exist, and the second term is 0 because there is no correlation surface in pipe from $(0,1,1)$ to $(0,1,2)$.
Although it is in general costly to express parity operator with first-order logic, we are only dealing with three or four terms, so the translation is simple.

\textit{All or No Orthogonal Surfaces at Non-$Y$ Cubes} (\autoref{fig:functional-constraints}c, generalizes \autoref{fig:corrsurf}b-c).
The correlation surface pieces orthogonal to the normal direction at a cube should all be present or all missing.
Let us consider $(1,0,1)$ with normal direction $I$:
\begin{equation}
    \begin{split}
        & \big[\ \overline{\texttt{YCube}_{1,0,1}}\ \wedge \overline{\texttt{ExistI}_{0,0,1}}\  \wedge \overline{\texttt{ExistI}_{1,0,1}}\ \big] \Rightarrow \\
        & \ \big\{ \big[\ \overline{\texttt{ExistJ}_{1,0,1}}\ \vee \texttt{CorrJK}_{1,1,0,1} \big] \wedge \big[\ \overline{\texttt{ExistK}_{1,0,0}} \\
        & \ \ \vee \texttt{CorrKJ}_{1,1,0,0} \big] \wedge \big[\ \overline{\texttt{ExistK}_{1,0,1}}\ \vee \texttt{CorrKJ}_{1,1,0,1} \big] \big\} \bigvee \\
        & \  \big\{ \big[ \ \overline{\texttt{ExistJ}_{1,0,1}}\ \vee \overline{\texttt{CorrJK}_{1,1,0,1}} \ \big] \wedge \big[ \ \overline{\texttt{ExistK}_{1,0,0}} \\
        & \ \ \vee \overline{\texttt{CorrKJ}_{1,1,0,0}} \ \big] \wedge \big[ \ \overline{\texttt{ExistK}_{1,0,1}}\ \vee \overline{\texttt{CorrKJ}_{1,1,0,1}} \ \big] \big\},
    \end{split}
\end{equation}
where we have two options on the right hand side corresponding to either correlation surface pieces are all present or all missing.
In each option, there are three terms corresponding to the three possible pipes connecting $(1,0,1)$.
In each term, if the \texttt{Exist} variable is 0, the term trivially evaluates to 1.
This corresponds to the fact that if a pipe does not exist, we do not need to consider correlation surface variables inside it.

\section{Software Implementation} \label{sec:implementation}

\begin{figure}[t]
    \centering
    \includegraphics[width=\linewidth]{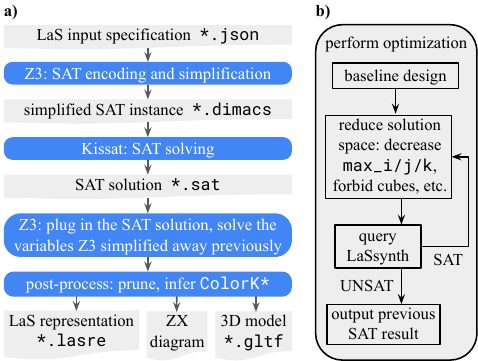}
    \vspace{-20pt}
    \caption{Software implementation.
    \textbf{a)}~Workflow of our LaS synthesizer, LaSsynth.
    \textbf{b)}~Performing LaS optimizations with LaSsynth.
    }
    \label{fig:implementation}
    % \vspace{-15pt}
\end{figure}

The structure of our LaS synthesizer, LaSsynth, is exhibited in \autoref{fig:implementation}a.
Given an input file following the specification in \autoref{fig:subroutine}b, we add variables and constraints in the formulation to an SMT (satisfiability modulo theories) model in Z3 SMT solver~\cite{tacas08-demoura-bjorner-z3-smt-solver}.
The model can be solved directly in Z3, but in our experience, the internal solver does not offer the best performance.
Thus, while keeping the option of solving the model directly, we support using Z3 just to simplify the model and transform it to a SAT instance stored in the standard SAT format--DIMACS.
Then, we use another SAT solver, Kissat~\cite{BiereFleury-SAT-Competition-2022-solvers}, to solve it.
Since DIMACS is the standard format, it is straightforward to port to any SAT solver on the market with minimal code changes.
We chose to still keep Z3 in the loop because some of its simplification `tactics', e.g., \texttt{simplify} and \texttt{propagate-values} make a big difference to later SAT solving in our experience.
After Kissat solves the SAT instance, we return to Z3, set the variable values according to the SAT result, and let Z3 solve the model again.
This second solving is negligible compared to the first one since Z3 is just re-deriving the variables it simplified away previously.
In our experiment, it never goes on more than 1s.

At this point, we have derived the values for all the variables in the formulation.
However, given how we have formulated the problem, the SAT solver has no preference for empty space, so the solution found may contain structures that are valid but unnecessary.
As post-processing, if a cube is not connected to any ports, it can be pruned away because it has no effect on the functionality of the LaS.
Usually, what the pruning removes are pipe ``donuts'' isolated from all other pipes.
Another post-processing step is coloring the $K$-pipes, i.e., deriving two arrays of additional values, \texttt{ColorKP} and \texttt{ColorKM}, for the color of $K$-pipe at the upper and lower ends.
If a $K$-pipe end is ``dangling'', it must be a port, then its color is given in the input.
If it touches any $I$- or $J$-pipes, its color can be inferred following the color matching constraints.
Finally, if it is in a $K$ passthrough, we can always add domain walls to legalize any $K$-pipe with different colors at two ends.

LaSsynth has three kinds of output.
All the variable assignments constitute our textual LaS representation, LaSre.
The second output are 3D modeling files in \texttt{gltf} format.
In fact, we will present pictures rendered using these generated files in the evaluation section.
We can also generate the 3D model with a specific correlation surface like \autoref{fig:functional-vars}.
Finally, LaSsynth can induce a ZX diagram of the LaS.
It uses \href{https://github.com/quantumlib/Stim/tree/main/glue/zx}{Stim ZX}~\cite{gidney2021stim} to derive the stabilizers of this ZX diagram, and compare these stabilizers with the ones given in the input for verification.

In summary, LaSsynth consumes a specification file and either asserts it is unsatisfiable or produces a LaS satisfying the input.
To perform \textit{optimization}, we need to query LaSsynth multiple times as illustrated by \autoref{fig:implementation}b.
If there is a known LaS design, we can treat it as the baseline and revise the specification to reduce the solution space in search for better designs.
The simplest example is shrinking the allowed volume by decreasing \texttt{max\_i/j/k}.
We also provide the interface to forbid certain cubes by setting all the \texttt{Exist} variables of its pipes to 0, which can be useful if the user is enforcing more complex shapes.
The user can also define more general techniques because we provide the interface to set the values of an arbitrary variable in the SMT model to the user-provided value.
After the solution space reduces, we query LaSsynth again, until there is no solution, by which we know that the optimal solution is the last satisfying solution.
This approach is descending in the sense that we start from a higher-volume solution and gradually find lower-volume solutions.
A drawback of this approach is that sometimes the baseline design is too bad, and the SAT solving takes a very long time given the unnecessarily large dimensions.
Thus, sometimes it also helps to take an ascending approach where we start at unsatisfiable settings, e.g., a very small \texttt{max\_i/j/k}, and grow the solution space until LaSsynth returns a solution.

To exploit certain flexibility in the problem, we may also query LaSsynth many times to search for designs, but not necessarily changing the size of the solution space.
For example, in the $T$-factory later discussed, many ports are functionally symmetric, but when we lay them out in the 3D space, some symmetries are broken.
This means, even with the same allowed volume, some permutations of the ports may be satisfiable while others are not.
In this case, we can generate a specification file for promising permutations and run many LaSsynth jobs in parallel.
Other than the order of the ports, the location of the ports can also be flexible sometimes.
Again, we can run many LaSsynth jobs in parallel, one for each possibility.
We have not implemented any general interface to perform explorations under these flexibilities because it greatly depends on the properties of the problem the user has.

\begin{figure*}[t]
    \centering
    \includegraphics[width=\linewidth]{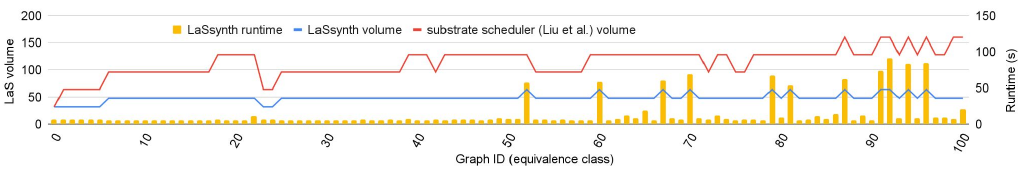}
    \vspace{-22pt}
    \caption{ LaS volume and LaSsynth runtime for 8-qubit graph state generation. Each graph represents a local-unitary equivalence class of graph states.}
    \label{fig:graph-result}
    % \vspace{-8pt}
    % updated
\end{figure*}

\section{Evaluation} \label{sec:evaluation}

We implemented LaSsynth (code provided \cite{tan_2024_11051465}) in Python3 with a dependency on Z3 4.12.1.0~\cite{tacas08-demoura-bjorner-z3-smt-solver}.
Installing Kissat 3.1.0~\cite{BiereFleury-SAT-Competition-2022-solvers} is required for the recommended SAT solving path.
Installing Stim ZX~\cite{gidney2021stim} is optional for verifying the stabilizers of LaS.
The following evaluations are done on a Linux server with two AMD EPYC 7V13 Processor and 512GB DRAM.

\subsection{Summary of Results} \label{ssec:summary_results}

Generic graph state generation provides a representative scenario for the intended use of LaSsynth.
In comparison with the compiler from Ref.~\cite{liu2023substrate}, tailored for a (logical) 2-lane architecture with careful qubit initialization and optimal gate scheduling, LaSsynth outperforms this baseline by 56\% on average across a comprehensive 8-qubit graph state benchmark set, as shown in \autoref{fig:graph-result}.
Our advantage lies in the smaller footprint per logical qubit, coupled with the ability to establish more intricate connectivity for accessing them.

We can utilize LaSsynth to construct non-Clifford LaS by inputting non-Clifford resource from ports.
For instance, in the majority gate, an important LaS in Shor's algorithm, three ports consume a $|$CCZ$\rangle$.
LaSsynth reduces volume by 40\% compared to the design in Ref.~\cite{gidney2019flexible}.
The corresponding ZX diagram is challenging for human understanding because of the creative use of generalized Hopf rule (see \autoref{fig:majority}).
Our ZX calculus verification reaffirms the correctness of our result and actually reveals the error of the design in Ref.~\cite{gidney2019flexible}.

We leverage LaSsynth to optimize $T$-factories, the dominating cost in FTQC.
There are some nuanced considerations at the non-Clifford input ports because of nondeterministic state injections.
Although we choose to get around these intricacies with very basic techniques, LaSsynth still discovers a 15-to-1 $T$-factory, showcased in \autoref{fig:t-factory}, 8\% smaller than state-of-the-art design~\cite{fowler2019low, Gidney2019efficientmagicstate}.
If neglecting state injection delays, LaSsynth discovers a design, as depicted in \autoref{fig:litinski}b, 18\% smaller than the state-of-the-art in this setting~\cite{Litinski2019gameofsurfacecodes}, using a smaller footprint while maintaining the same depth.

The primary use case for LaSsynth is optimizing critical subroutines.
Despite potential exponential scaling of the internal SAT solving, it consistently outperforms human expert designs at the scale of realistically significant subroutines, as evidenced by the presented results.
In essence, our advantage lies in more flexibility of allocating, moving, and recycling code patches.
As pipe diagrams, the human designs usually let the qubits stay put, i.e., forming ``pillars'' that vertically goes through the LaS, and lattice surgery is performed by unit-time horizontal crossbars connecting to the pillars.
In contrast, the vertical pipes in our generated LaS can terminate and begin at will, so the solver explores a much larger design space, e.g., ancillas may not be immediately recycled, but squeezed and moved around to interact with other qubits.

\subsection{Methodology of Graph State Generation Evaluation}

\begin{figure}[t]
    \centering
    \includegraphics[width=\linewidth]{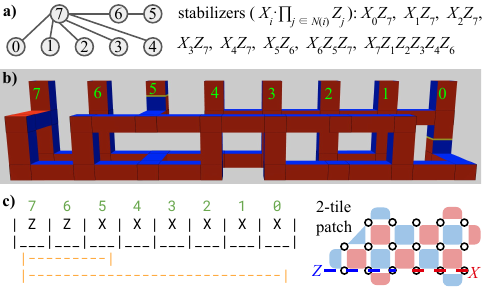}
    \vspace{-20pt}
    \caption{\textbf{a)}~Graph state.
    \textbf{b)}~8$\times$2$\times$2 LaS generating the graph state found by LaSsynth.
    \textbf{c)}~8$\times$4$\times$2 solution found by the baseline~\cite{liu2023substrate}.
    They require 2-tile patches instead of 1-tile patches in our case.
    Letters indicate initial basis of patches.
    Orange intervals indicate two layers of parity measurements.
    }
    \label{fig:graph-state}
    % \vspace{-15pt}
\end{figure}

\textit{Graph state.} 
The stabilizers of a graph state with underlying graph $G=(V,E)$ are generated by $\{ X_i\cdot\prod_j Z_j\ |\ \forall i\in V,\ (i,j)\in E \}$, i.e., $X$ on a node and $Z$ on all its neighbors in $G$, e.g., \autoref{fig:graph-state}a.
These states have many applications in FTQC~\cite{book06-graph-state, vijayan2022compilation}.
In fact, any state defined by Pauli string stabilizers is equivalent to a graph state up to single-qubit unitaries, so generic graph state generation serves as the average case of LaS, which is the target use case of our tool.

\textit{Hardware assumption.}
Aligning with the baseline~\cite{liu2023substrate}, we adopt a (logical) 2-lane architecture.
There are 2 lanes of surface code tiles available as workspace, each tile is $d\times d$ physical qubits.
In our specification, this means limiting $\texttt{max\_j}=2$.
\autoref{fig:graph-state}b provides an illustration, where the back lane is for (logical) qubit output, and the front lane is ancillary.

\textit{Baseline approach.}
Liu et al. developed a compiler~\cite{liu2023substrate} based on Ref.~\cite{Litinski2019gameofsurfacecodes}: initializing logical qubits in selective basis and then performing multi-qubit parity measurements using lattice surgery.
They observed that selecting the initialization basis is a Maximum Independent Set problem (MIS).
Given the initialization in \autoref{fig:graph-state}c, only the last two stabilizers in \autoref{fig:graph-state}a require measurement.
MIS is NP-hard, so, to be fair, \textit{we gave this compiler the same amount of time as LaSsynth spends.}
In their setting, to enable measurements in both the $X$ and $Z$ bases, the qubits must expose both types of boundaries to the ancilla lane, necessitating 2-tile patches (on the right side of \autoref{fig:graph-state}c), pushing footprint of their LaS to 16$\times$2=32.

\textit{Comprehensive benchmark set.}
There are $2^{n(n-1)/2}$ graphs for $n$ nodes, but many of them are equivalent up to 1-Q Cliffords.
Our benchmarks are 101 graphs from a \href{http://www.ii.uib.no/~larsed/entanglement/}{database}~\cite{graph-equivalence} representing \textit{all} the equivalence classes of 8-qubit graphs.

\textit{Our approach.}
Because of the 2-lane assumption, we specify 8$\times$2 footprint along with the graph state stabilizers to LaSsynth.
We initiate the search at a depth 3 and iteratively adjust it based on the response—increasing depth if UNSAT or decreasing it if SAT—to determine the optimal depth.
The resulting LaS volume is then 8$\times$2$\times$ the optimal depth.

% \textit{Data analysis.}
% In \autoref{fig:graph-result}, we present a comparison of the LaS volumes produced by our approach and the baseline.
% On average, LaSsynth achieves a 56\% reduction in volume.
% In \autoref{table:graph}, we categorize the graphs based on the number of edges (7-13), revealing that the advantage of LaSsynth becomes more pronounced with higher edge occupancy.
% The baseline LaS volume consistently increases, while our LaS volume peaks at 11 edges, aligning with a theoretical result indicating that graph states with ``moderate'' connectivity are the most ``complex''~\cite{graph-state-complexity}.
% Notably, the baseline can explore changing the order of qubits at the output (`remap' in the table), a feature not supported by LaSsynth.
% Despite this limitation, our results exhibit significant improvements.

\textit{Source of advantage.}
The baseline approach relies on 2-tile patches, doubling the required footprint and placing it at a disadvantage.
This necessity stems from the limitation of 1-tile patches, where only one basis is exposed to the ancilla at a time, posing a challenge for human intuition to access all required bases effectively.
LaSsynth overcomes this limitation by employing domain walls (depicted as yellow rings) and intricate connectivity, as seen in \autoref{fig:graph-state}b.
In contrast, the baseline solutions would consist only of horizontal bars, i.e., parity measurements, connecting to some of the 8 qubits.

% \begin{table}
%   \centering
%   \caption{8-Qubit graph state generation subroutine volume}
%   \label{table:graph}
%       %\vspace{-5pt}
%   \begin{tabular}{|c|c|c|c|c|c|}
%     \hline
%     Edges & ID & Ours & Liu et al.~\cite{liu2023substrate} & \cite{liu2023substrate} remap & Impr. \\
%     \hline
%     7 & 0-22 & 43.8 & 93.2 & 64.0 & 53.0\%\\
%     8 & 23-54 & 47.5 & 107 & 90.0 & 55.6\%\\
%     9 & 55-74 & 50.4 & 120 & 97.6 & 58.0\%\\
%     10 & 75-87 & 51.7 & 126 & 111 & 58.8\%\\
%     11 & 88-94 & 54.9 & 142 & 128 & 61.3\%\\
%     12 & 95-98 & 53.3 & 139 & 139 & 61.5\%\\
%     13 & 98-100 & 48.0 & 149 & 149 & 67.9\%\\
%     \hline
%   \end{tabular}
%       %\vspace{-10pt}
% \end{table}

\subsection{Methodology of Non-Clifford Subroutine Evaluation}

\textit{Handling non-Cliffordness.}
The functionality of LaS is specified by stabilizers, implying it can always be implemented using Clifford gates and Pauli measurements.
However, for FTQC, non-Cliffordness like $|T\rangle$ is required, and lattice surgery alone cannot fully implement them.
Typically, non-Clifford resources are generated in dedicated regions on the quantum chip and routed to specific places.
By considering non-Clifford resources as input to certain ports, the remaining portion constitutes a LaS specified by stabilizers.
Thus, our LaS definition is enough for general FTQC subroutines.
Additionally, this definition is not limited to specific types of non-Cliffordness; for example, we accommodate the majority gate in Ref.~\cite{gidney2019flexible}, which consumes a $|\text{CCZ}\rangle$ instead of $|T\rangle$.

\begin{figure}[t]
    \centering
    \includegraphics[width=\linewidth]{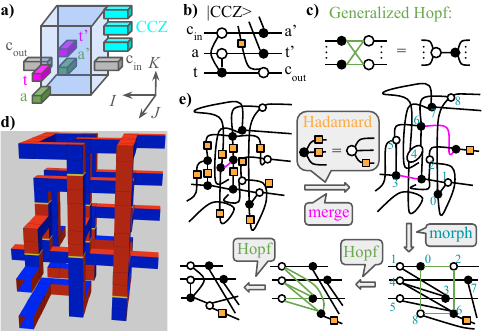}
    \vspace{-20pt}
    \caption{Majority gate optimization.
    \textbf{a)}~Port requirements. Ref.~\cite{gidney2019flexible} provides a 3$\times$5$\times$5 design.
    \textbf{b)}~ZX diagram of the gate.
    \textbf{c)}~Hopf rule in ZX calculus.
    \textbf{d)}~3$\times$3$\times$5 design found by LaSsynth.
    \textbf{e)}~ZX calculus proof of the LaS.
    }
    \label{fig:majority}
    % \vspace{-15pt}
\end{figure}

\textit{Port requirements of the majority gate.}
To align with the use case in Ref.~\cite{gidney2019flexible} (see \autoref{fig:majority}a), some port location requirements must be met.
Specifically, $t$ and $t'$ must be at the same height, so do $a$ and $a'$, and $c_\text{in}$ and $c_\text{out}$.
The routed $|\text{CCZ}\rangle$ necessitates three additional vertically aligned ports above them.
These constraints imply $\texttt{max\_j}\ge 3$ and $\texttt{max\_k}\ge 5$, leaving the only dimension that can shrink as $I$.

\textit{Importance of verification.}
To verify a design, LaSsynth extracts a ZX diagram, and leverages Stim ZX to derive its stabilizers.
When we read off the 5$\times$3$\times$5 design in Ref.~\cite{gidney2019flexible} to a LaSre and give it to LaSsynth, verification fails, underscoring the susceptibility of human designed LaS to errors and the practical challenges of manual verification, even for experts.

\textit{Trying to interpret the generated design.}
LaSsynth derived a LaS (\autoref{fig:majority}d) 40\% smaller than the baseline.
Mechanically following the constraints will not aid interpretation, so we use ZX calculus rewrite rules to prove that the generated LaS is equivalent to the ZX diagram of a majority gate (\autoref{fig:majority}b).
The initial diagram in \autoref{fig:majority}e is directly extracted from the pipes.
In the first step, we apply the Hadamard inversion rule (orange) and the spider-merge rule (magenta).
In the second step, we morph the diagram without altering connectivity (numbers annotate corresponding spiders).
Next, we apply the (generalized) Hopf rule (green), illustrated in \autoref{fig:majority}c: when $Z$- and $X$-spiders are connected in a complete bipartite manner, two new spiders can replace them.
Applying this rule twice reconstructs the ZX diagram in \autoref{fig:majority}b, completing our proof.
The two Hopf rules significantly alter the connectivity of the diagram, producing a compact yet intricate LaS.

\subsection{Methodology of $T$-Factory Evaluation}

\textit{Distillation Factory Motivation.}
The majority gate consumes non-Cliffordness, but where do those come from?
A prevalent solution are magic state distillation factories (usually for $|T\rangle$ or $|\text{CCZ}\rangle$) that consume noisy magic states and produce higher-quality magic states.
Our LaS model can support such factories: a first-level factory involves physical magic state injections at certain ports, while a higher-level factory takes already distilled magic states to the ports; the remaining part can still be specified by stabilizers.
Distilling an injected $|T\rangle$ to a usable error rate entails thousands of operations, making non-Clifford gates a significant FTQC cost.
Reducing distillation factory sizes directly reduces this dominant cost.

\begin{figure}[t]
    \centering
    \includegraphics[width=\linewidth]{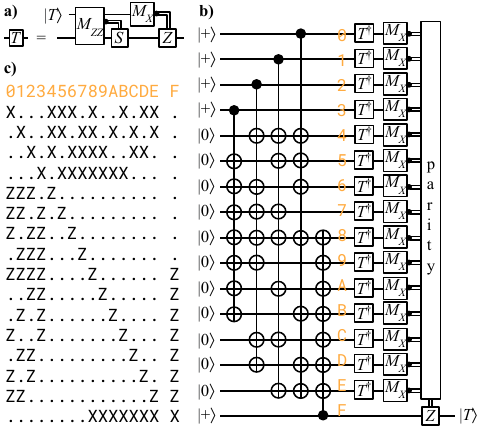}
    \vspace{-20pt}
    \caption{Fault-tolerant $T$.
    \textbf{a)}~Implementing a $T$ gate by consuming a magic state.
    \textbf{b)}~Quantum circuit for the 15-to-1 $T$-factory~\cite{fowler2019low}.
    \textbf{c)}~Stabilizers of the underlying [[15,1,3]] error correcting code.
    }
    \label{fig:t-gate}
    % \vspace{-15pt}
\end{figure}

\textit{The 15-to-1 T-factory baseline.}
The 15-to-1 $T$-factory is one of the most realistic choices for early FTQC, visualized in \autoref{fig:t-gate}b in the circuit model.
The 15 $T^\dagger$'s and the final output $|T\rangle$ correspond to 16 non-Clifford ports in our LaS.
The remaining portion is specified by stabilizers at the locations marked with orange labels, derived in \autoref{fig:t-gate}c.
A baseline design, manually optimized by experts in Refs.~\cite{fowler2019low, Gidney2019efficientmagicstate}, initializes logical qubits in $|0\rangle$ with unit depth, an eigenstate of the $Z$ stabilizers in \autoref{fig:t-gate}c.
It then measures the 5 8-body $X$ stabilizers in \autoref{fig:t-gate}c, each taking unit depth.
A layer of possible \textit{fixups} (see next paragraph) is appended, resulting in a total LaS depth of 6.5.
When used repeatedly, one unit of latency is hidden through interleaving, and the baseline factory averages a depth of 5.5.
With a footprint of 8$\times$4=32, the baseline factory has an average volume of 32$\times$5.5=176.

\begin{figure}[t]
    \centering
    \includegraphics[width=\linewidth]{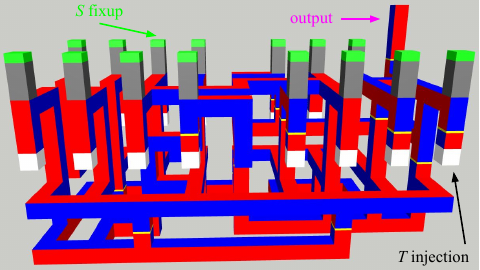}
    \vspace{-20pt}
    \caption{Pipe diagram of a 15-to-1 $T$-factory with $9\times 4 \times 4.5$ spacetime volume.
    White boxes are magic state injections.
    An $S$ fixup consists of conditional $K$-pipes (gray) and $Y$ cubes (green).
    }
    \label{fig:t-factory}
    % \vspace{-5pt}
\end{figure}

\textit{Fixups.}
When injecting $|T\rangle$ in the surface code, half the time yields $|T\rangle$, and the other half yields $|T^\dagger\rangle$~\cite{fowler2019low}.
Consequently, we may need to apply an $S$ gate via $Y$ cubes based on the injection result.
These dynamic cubes are not included in the formulation since they depend on \textit{runtime} information.
We need to reserve space in the LaS for these cubes.
How much space and where is necessary to accommodate any of the $2^{15}$ injection outcomes is an involved topic.
For simplicity, we adopt a straightforward technique illustrated in \autoref{fig:t-factory}: each injection connects to a $K$-pipe and then bends inward to the rest of the pipe diagram.
Fixups for each injection can be attached where the pipe bends.
We do not include the fixups to the specification and append the fixup layer after synthesis.

\textit{T-factory optimization results.}
The baseline design actually employs \textit{half-distance rotation}, not available in our formulation.
Despite this disadvantage, by iteratively calling LaSsynth with shrinking volume, we obtained solutions with volumes of 7$\times$5$\times$5=175 and 6$\times$6$\times$4.5=9$\times$4$\times$4.5=162.
In \autoref{fig:t-factory}, we showcase one of the best designs, 8\% smaller than the baseline.
We verified \href{https://algassert.com/zigxag\#0,0,O;0,-6,O;0,-12,out;0,-18,O;-1,-1,O;-1,-7,@;-1,-13,O;-1,-19,O;-2,-2,O;-2,-8,O;-2,-14,O;-2,-20,O;-3,-3,O;-3,-9,O;-3,-15,out;-3,-21,O;5,0,@;5,-6,O;5,-12,out;5,-18,O;4,-1,@;4,-7,O;4,-13,O;4,-19,O;3,-2,O;3,-8,@;3,-14,O;3,-20,O;2,-3,O;2,-9,O;2,-15,out;2,-21,O;10,0,O;10,-6,@;10,-12,out;10,-18,O;9,-1,O;9,-7,@;9,-13,O;9,-19,O;8,-2,O;8,-8,@;8,-14,O;8,-20,@;7,-3,O;7,-9,O;7,-15,out;7,-21,O;15,0,O;15,-6,O;15,-12,out;15,-18,O;14,-7,O;14,-13,O;14,-19,@;13,-2,O;13,-8,@;13,-14,O;13,-20,O;12,-3,O;12,-9,O;12,-15,out;12,-21,O;20,0,O;20,-6,O;20,-12,O;20,-18,O;19,-1,O;19,-7,@;19,-13,O;19,-19,O;18,-2,O;18,-8,O;18,-14,@;18,-20,O;17,-3,O;17,-9,O;17,-15,O;17,-21,O;25,0,O;25,-6,O;25,-12,out;25,-18,O;24,-1,O;24,-7,@;24,-13,O;24,-19,O;23,-2,O;23,-8,O;23,-14,O;23,-20,@;22,-3,O;22,-9,O;22,-15,out;22,-21,O;30,0,O;30,-6,@;30,-12,out;30,-18,O;29,-1,O;29,-7,@;29,-13,O;29,-19,@;28,-2,O;28,-8,@;28,-14,O;28,-20,O;27,-3,O;27,-9,O;27,-15,out;27,-21,O;35,-6,O;35,-12,out;35,-18,O;34,-1,O;34,-7,@;34,-13,O;34,-19,O;33,-2,O;33,-8,O;33,-14,O;33,-20,@;32,-3,O;32,-9,O;32,-15,out;32,-21,O;40,-6,O;40,-12,out;40,-18,O;39,-1,O;39,-7,@;39,-13,O;39,-19,O;38,-14,@;38,-20,@;37,-15,O;37,-21,O;37,-27,out:0,0,5,0,-;0,0,-1,-1,-;0,-6,5,-6,-;0,-6,-1,-7,-;0,-12,0,-18,-;0,-18,-1,-19,-;-1,-1,-2,-2,-;-1,-1,-1,-7,h;-1,-7,-1,-13,-;-1,-13,-1,-19,-;-2,-2,-3,-3,-;-2,-8,-3,-9,-;-2,-8,-2,-14,-;-2,-14,-2,-20,-;-2,-20,-3,-21,-;-3,-3,2,-3,-;-3,-9,2,-9,-;-3,-15,-3,-21,-;5,0,10,0,-;5,0,4,-1,-;5,-6,10,-6,-;5,-6,4,-7,-;5,-12,5,-18,-;5,-18,4,-19,-;4,-1,9,-1,-;4,-1,3,-2,-;4,-7,4,-13,-;4,-13,4,-19,-;3,-2,3,-8,h;3,-8,2,-9,-;3,-8,3,-14,-;3,-14,3,-20,-;3,-20,2,-21,-;2,-3,7,-3,-;2,-9,7,-9,-;2,-15,2,-21,-;10,0,15,0,-;10,0,10,-6,h;10,-6,15,-6,-;10,-12,10,-18,-;10,-18,9,-19,-;9,-1,9,-7,h;9,-7,14,-7,-;9,-7,9,-13,-;9,-13,14,-13,-;9,-19,14,-19,-;9,-19,8,-20,-;8,-2,13,-2,-;8,-2,8,-8,-;8,-8,7,-9,-;8,-8,8,-14,-;8,-14,8,-20,-;8,-20,7,-21,-;7,-3,12,-3,-;7,-9,12,-9,-;7,-15,7,-21,-;15,0,20,0,-;15,-6,20,-6,-;15,-6,14,-7,-;15,-12,15,-18,h;15,-18,20,-18,-;14,-7,13,-8,-;14,-13,14,-19,-;14,-19,19,-19,-;13,-2,18,-2,-;13,-2,13,-8,-;13,-8,12,-9,-;13,-8,13,-14,-;13,-14,13,-20,h;13,-20,12,-21,-;12,-3,17,-3,-;12,-15,12,-21,h;20,0,25,0,-;20,-6,25,-6,-;20,-12,19,-13,-;20,-12,20,-18,-;19,-1,24,-1,-;19,-1,19,-7,-;19,-7,18,-8,-;19,-7,19,-13,h;19,-13,18,-14,-;19,-19,18,-20,-;18,-2,23,-2,-;18,-2,18,-8,-;18,-14,23,-14,-;18,-14,17,-15,-;18,-20,23,-20,-;17,-3,17,-9,h;17,-9,22,-9,-;17,-15,17,-21,-;17,-21,22,-21,-;25,0,30,0,-;25,-6,30,-6,-;25,-6,24,-7,-;25,-12,25,-18,h;25,-18,24,-19,-;24,-1,29,-1,-;24,-7,23,-8,-;24,-7,24,-13,-;24,-13,24,-19,h;23,-2,22,-3,-;23,-8,22,-9,-;23,-14,28,-14,-;23,-14,23,-20,h;23,-20,28,-20,-;22,-3,27,-3,-;22,-9,27,-9,-;22,-15,22,-21,h;30,0,30,-6,h;30,-6,35,-6,-;30,-12,30,-18,-;30,-18,29,-19,-;29,-1,34,-1,-;29,-1,29,-7,-;29,-7,28,-8,-;29,-7,29,-13,-;29,-13,29,-19,-;29,-19,28,-20,-;28,-2,33,-2,-;28,-2,28,-8,-;28,-8,27,-9,-;28,-14,33,-14,-;28,-20,27,-21,-;27,-3,32,-3,-;27,-9,32,-9,-;27,-15,27,-21,-;35,-6,40,-6,-;35,-6,34,-7,-;35,-12,35,-18,h;35,-18,34,-19,-;34,-1,39,-1,-;34,-1,34,-7,-;34,-7,33,-8,-;34,-7,34,-13,-;34,-13,34,-19,h;34,-19,33,-20,-;33,-2,32,-3,-;33,-8,32,-9,-;33,-14,38,-14,-;33,-20,38,-20,-;33,-20,32,-21,-;32,-15,32,-21,h;40,-6,39,-7,-;40,-12,40,-18,h;40,-18,39,-19,-;39,-1,39,-7,-;39,-7,39,-13,h;39,-13,38,-14,-;39,-13,39,-19,-;39,-19,38,-20,-;38,-14,37,-15,-;38,-20,37,-21,-;37,-15,37,-21,-;37,-21,37,-27,h}{its ZX diagram} using Stim ZX.
Notably, by avoiding half-distance rotations present in the baseline, this design opens opportunities for using half-distance elsewhere.
Thus, it is of interest to quantum error correction experts to explore how to reduce the distance in regions of this design, potentially achieving further improvements.

\begin{figure}[t]
    \centering
    \includegraphics[width=\linewidth]{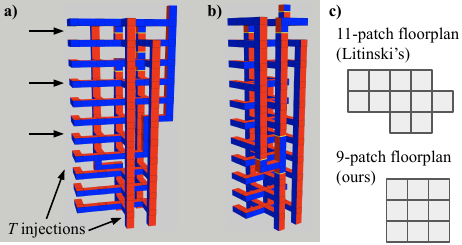}
    \vspace{-20pt}
    \caption{$T$-factory assuming no classical delay for injections.
    \textbf{a)}~A generated design with volume 121, same as Ref.~\cite{Litinski2019gameofsurfacecodes}.
    \textbf{b)}~Optimized design with 3$\times$3$\times$11 volume.
    \textbf{c)}~Floorplan comparison of the two designs.
    }
    \label{fig:litinski}
    % \vspace{-15pt}
\end{figure}

\textit{T-factory assuming no classical delay.}
Much of the preceding discussion delves into fixup details.
Ignoring classical injection delay, Ref.~\cite{Litinski2019gameofsurfacecodes} presents a 121-volume factory design utilizing 11 patches (\autoref{fig:litinski}c top) and a depth of 11, with four injections from the bottom and the remainder from the side (\autoref{fig:litinski}a).
Under the same assumption, LaSsynth derives a design (\autoref{fig:litinski}b) with volume 3$\times$3$\times$11=99, achieving an 18\% reduction by using a smaller footprint (\autoref{fig:litinski}c bottom).

\subsection{Observations on Runtime}\label{ssec:obs_runtime}

\textit{Scalability metrics.}
The runtime of LaSsynth may scale unfavorably due to its dependence on SAT solving.
However, its target use case is frequently used subroutines rather than entire algorithms, and the presented results demonstrate its effectiveness in solving significant and realistic subroutines.
Some runtimes are recorded in \autoref{table:runtime-others}, where the formulation yields a scaling factor, $Vn_\text{stab}$, i.e., the volume times the number of stabilizers.
It is worth noting that this column may appear slightly larger than expected because of padding layers on boundaries to support ports or non-rectangular floorplans.
While the 121-factory has a larger $Vn_\text{stab}$ than the 162-factory, it is a simpler problem to solve.
A better indicator is the number of variables and clauses of the generated SAT CNF.

\textit{Random seed: more is different.}
There is potential for portfolio-based SAT solving~\cite{schreiber2021-mallob}.
We employed 10 random seeds on the same CNF, and the standard deviations of the runtimes are presented in the last column of \autoref{table:runtime-others}.
Notably, for the 162-factory, the runtime difference can be as much as 26 times, indicating that multiple SAT solvers with different seeds may significantly expedite finding a solution.

\textit{To UNSAT or not, it is a question.}
Unsatisfiable specifications took longer than satisfiable ones.
In \autoref{fig:graph-result}, it is noticeable that long runtimes consistently accompany a `spike' in volume.
These instances represent cases where LaSsynth initiates with a depth of 3, requires a relatively lengthy period to determine unsatisfiability, and subsequently discovers a solution with depth 4.
Generally, this is the price to pay for the optimality guarantee.
For scenarios where optimal solutions are not strictly necessary, prioritizing satisfiability initially—such as with incomplete approaches like MaxSAT, or iterative solving, retaining learned information—can be more beneficial in obtaining good designs efficiently.

\begin{table}[t]
    \centering
    \caption{Size and runtime for presented non-Clifford designs}
    \label{table:runtime-others}
    \vspace{-5pt}
    \begin{tabular}{|c|c|c|c|c|c|}
      \hline
      Name &  $Vn_\text{stab}$ & Variables & Clauses & Min. Time (s) & SD \\
      \hline
      Majority & 720 & 8173 & 56851 & 9.02 & 0.33 \\
      99-factory  & 1728 & 33650 & 248974 & 20.6 & 0.61 \\
      121-factory  & 2880 & 35657 & 267544 & 40.9 & 8.3\\
      162-factory  & 2304 & 43070 & 326305 & 469 & 4e3 \\
      \hline
    \end{tabular}
    % \vspace{-10pt}
  \end{table}

\section{Related Works} \label{sec:related}
Ref.~\cite{lattice-surgery-translation} first provided a compiler that takes in ICM (initialization, CNOT, and measurement) representation and translates the gates to lattice surgery operations.
Later on, researchers opt for a more efficient operation with lattice surgery, multi-qubit Pauli measurements.
Thus, a few works focused on implementing FTQC on a 2D grid of qubits with the multi-qubit-Pauli-based gate set~\cite{beverland2022assessing, temporal-lattice-surgery, fowler2019low, Litinski2019gameofsurfacecodes}.
Ref.~\cite{litinski2022active} still used this gate set but discussed the advantage of having non-local connectivity.
In terms of software, Ref.~\cite{paler2020opensurgery} provided an instruction set for this gate set and Ref.~\cite{watkins2023high} provided a compiler.
However, as we demonstrated above, it is beneficial to consider optimizations beyond this gate set.

Some previous works focused on improving specific components, not generic quantum circuit.
Since the magic state factories take up a lot of volume, they have become a natural target for such optimizations.
Ref.~\cite{Litinski2019magicstate} further developed the aforementioned technique of selectively reducing code distances, which can be applied in combination with the optimizations we present in this work.
Ref.~\cite{Gidney2019efficientmagicstate} considered the interplay of $T$-factories with $|\text{CCZ}\rangle$-factories, and presented improved factory designs.
Ref.~\cite{gidney2019flexible} provided further improvements on $|\text{CCZ}\rangle$-factories.
However, all these works are manual efforts instead of an automated synthesizer.

There is a similar line of works for defect based FTQC on surface codes~\cite{defect-surface-code}.
The compilation problem is formulated as routing FTQC components~\cite{paler2016synthesis}.
After a manual approach of bridge compression was proposed~\cite{fowler2013bridge}, researchers encoded the problem to integer linear programming~\cite{bridge-based-compression, layout-synthesis-defect}, which is another kind of mathematical programming than SAT.
Heuristic compilation approaches have been presented for optimizing communication~\cite{autobraid, surface-code-communication, qec-bandwidth}, or for $T$-factory~\cite{magic-state-unit}.
However, defect-based computation is phasing out because of higher overheads than lattice surgery~\cite{Litinski2019gameofsurfacecodes, fowler2019low}.

Ref.~\cite{shutty_decoding_2022} utilizes an SMT solver to synthesize fault-tolerant Clifford circuit in a bottom-up fashion like this work.
However, the gate set consists of CNOT, $X$, $Y$, $Z$, $S$, and $H$, quite different from generic lattice surgery operations in this work.

\section{Conclusion and Future Directions} \label{sec:conclusion}
In this work, we formulate the problem of synthesizing lattice-surgery subroutines, LaS.
We define LaSre, a novel representation of LaS, allowing much more flexibility for lattice surgery to happen than existing works.
We develop LaSsynth, a synthesizer for LaS utilizing SAT solvers, which can also be leveraged to optimize and verify existing LaS.
We demonstrate remarkable results in experimental evaluation, including $18\%$ spacetime volume reduction for 15-to-1 $T$ state distillation.
Our work opens up the new directions below.

\textit{Using LaSsynth More.}
A major benefit of having such a synthesizer is saving human efforts.
We ran LaSsynth on tens of thousands of instances during the preparation of this paper, which is an unimaginable amount to humans.
We can also perform architecture evaluations with the synthesizer by comparing the best results on different architectures.
Specifically, we can explore the performance of quasi-1D architectures, or very small footprint architectures (which needs more time to run circuits, but easier to fabricate in the near term).

\textit{Better SAT Solving.}
There is still an opportunity to accelerate the SAT solving by more clever encoding of constraints, solver preconditioning, iterative solving, or just swapping out the solver.
In addition, approximate methods like MaxSAT can be leveraged to trade synthesis speed with quality.

\textit{Automatic Exploration of Port Configurations.}
As mentioned in \autoref{sec:implementation}, there is an opportunity to devise better exploration approaches of port ordering and locations, rather than brute-force trying all the possibilities in parallel.

\textit{Supporting Other Operations.} 
In principle it is viable to support other operations than what listed in \autoref{fig:lattice-ops} by adding more variables and constraints, much like what we did for $Y$ cubes.
If a new operation is found to be very efficient, it is worthwhile to consider extending the formulation for it.
    
\textit{Theory on Attaching Magic State Injections to LaS.}
The current synergy between magic state injections and the rest of the pipeline is quite simple. Further optimization in the number and placement of fixups relative to the injections could lead to significant optimizations for the $T$-factory.
    
\textit{Integration into FTQC algorithm-level compiler.}
While our main focus in this work is strictly on subroutines, the developed LaSsynth can integrate into a full-algorithm level compiler.
This integration facilitates iterative subroutine optimization on top of algorithm-level compilation.

\section*{Contribution and Acknowledgment}
D.~B.~T. wrote the code and the initial draft of this paper as a student researcher at Google.
C.~G. and M.~Y.~N. played a supervisory role, guiding the project and giving design input.
D.~B.~T. is also grateful for discussions with Noah~Shutty and Sergio~Boixo at Google, and Prof.~Jason~Cong at UCLA.

%%%%%%% -- PAPER CONTENT ENDS -- %%%%%%%%
\clearpage

%%%%%%%%% -- BIB STYLE AND FILE -- %%%%%%%%
\bibliographystyle{myieeedoi}
\bibliography{refs}
%%%%%%%%%%%%%%%%%%%%%%%%%%%%%%%%%%%%
\end{document}